\PassOptionsToPackage{unicode}{hyperref}
\PassOptionsToPackage{hyphens}{url}
\PassOptionsToPackage{dvipsnames,svgnames,x11names}{xcolor}
\documentclass[
  letterpaper,
  DIV=11,
  numbers=noendperiod]{scrartcl}

\usepackage{amsmath,amssymb}
\usepackage{iftex}
\ifPDFTeX
  \usepackage[T1]{fontenc}
  \usepackage[utf8]{inputenc}
  \usepackage{textcomp} 
\else 
  \usepackage{unicode-math}
  \defaultfontfeatures{Scale=MatchLowercase}
  \defaultfontfeatures[\rmfamily]{Ligatures=TeX,Scale=1}
\fi
\usepackage{lmodern}
\ifPDFTeX\else  
\fi
\IfFileExists{upquote.sty}{\usepackage{upquote}}{}
\IfFileExists{microtype.sty}{
  \usepackage[]{microtype}
  \UseMicrotypeSet[protrusion]{basicmath} 
}{}
\makeatletter
\@ifundefined{KOMAClassName}{
  \IfFileExists{parskip.sty}{%
    \usepackage{parskip}
  }{
    \setlength{\parindent}{0pt}
    \setlength{\parskip}{6pt plus 2pt minus 1pt}}
}{
  \KOMAoptions{parskip=half}}
\makeatother
\usepackage{xcolor}
\ifx\paragraph\undefined\else
  \let\oldparagraph\paragraph
  \renewcommand{\paragraph}[1]{\oldparagraph{#1}\mbox{}}
\fi
\ifx\subparagraph\undefined\else
  \let\oldsubparagraph\subparagraph
  \renewcommand{\subparagraph}[1]{\oldsubparagraph{#1}\mbox{}}
\fi

\providecommand{\tightlist}{%
  \setlength{\itemsep}{0pt}\setlength{\parskip}{0pt}}\usepackage{longtable,booktabs,array}
\usepackage{calc} 
\usepackage{etoolbox}
\makeatletter
\patchcmd\longtable{\par}{\if@noskipsec\mbox{}\fi\par}{}{}
\makeatother
\IfFileExists{footnotehyper.sty}{\usepackage{footnotehyper}}{\usepackage{footnote}}
\makesavenoteenv{longtable}
\usepackage{graphicx}
\makeatletter
\def\maxwidth{\ifdim\Gin@nat@width>\linewidth\linewidth\else\Gin@nat@width\fi}
\def\maxheight{\ifdim\Gin@nat@height>\textheight\textheight\else\Gin@nat@height\fi}
\makeatother
\setkeys{Gin}{width=\maxwidth,height=\maxheight,keepaspectratio}
\makeatletter
\def\fps@figure{htbp}
\makeatother
\NewDocumentCommand\citeproctext{}{}

\makeatletter
 \let\@cite@ofmt\@firstofone
 \def\@biblabel#1{}
 \def\@cite#1#2{{#1\if@tempswa , #2\fi}}
\makeatother
\newlength{\cslhangindent}
\newlength{\csllabelwidth}
\newenvironment{CSLReferences}[2] 
 {\begin{list}{}{%
  \setlength{\itemindent}{0pt}
  \setlength{\leftmargin}{0pt}
  \setlength{\parsep}{0pt}
  \ifodd #1
   \setlength{\leftmargin}{\cslhangindent}
   \setlength{\itemindent}{-1\cslhangindent}
  \fi
  \setlength{\itemsep}{#2\baselineskip}}}
 {\end{list}}
\usepackage{calc}

\usepackage{booktabs}
\usepackage{longtable}
\usepackage{array}
\usepackage{multirow}
\usepackage{wrapfig}
\usepackage{float}
\usepackage{colortbl}
\usepackage{pdflscape}
\usepackage{tabu}
\usepackage{threeparttable}
\usepackage{threeparttablex}
\usepackage[normalem]{ulem}
\usepackage{makecell}
\usepackage{xcolor}
\usepackage{siunitx}

  \newcolumntype{d}{S[
    input-open-uncertainty=,
    input-close-uncertainty=,
    parse-numbers = false,
    table-align-text-pre=false,
    table-align-text-post=false
  ]}
  
\KOMAoption{captions}{tableheading}
\usepackage{listings}
\makeatletter
\@ifpackageloaded{caption}{}{\usepackage{caption}}
\AtBeginDocument{%
\ifdefined\contentsname
  \renewcommand*\contentsname{Table of contents}
\else
  \newcommand\contentsname{Table of contents}
\fi
\ifdefined\listfigurename
  \renewcommand*\listfigurename{List of Figures}
\else
  \newcommand\listfigurename{List of Figures}
\fi
\ifdefined\listtablename
  \renewcommand*\listtablename{List of Tables}
\else
  \newcommand\listtablename{List of Tables}
\fi
\ifdefined\figurename
  \renewcommand*\figurename{Figure}
\else
  \newcommand\figurename{Figure}
\fi
\ifdefined\tablename
  \renewcommand*\tablename{Table}
\else
  \newcommand\tablename{Table}
\fi
}
\@ifpackageloaded{float}{}{\usepackage{float}}
\floatstyle{ruled}
\@ifundefined{c@chapter}{\newfloat{codelisting}{h}{lop}}{\newfloat{codelisting}{h}{lop}[chapter]}
\floatname{codelisting}{Listing}

\makeatother
\makeatletter
\@ifpackageloaded{caption}{}{\usepackage{caption}}
\@ifpackageloaded{subcaption}{}{\usepackage{subcaption}}
\makeatother
\ifLuaTeX
  \usepackage{selnolig}  
\fi
\usepackage{bookmark}

\IfFileExists{xurl.sty}{\usepackage{xurl}}{} 
\hypersetup{
  pdftitle={Evaluating the Decency and Consistency of Data Validation Tests Generated by LLMs},
  pdfauthor={Rohan Alexander; Lindsay Katz; Callandra Moore; Michaela Drouillard; Michael Wing-Cheung Wong; Zane Schwartz},
  colorlinks=true,
  linkcolor={blue},
  filecolor={Maroon},
  citecolor={Blue},
  urlcolor={Blue},
  pdfcreator={LaTeX via pandoc}}

\title{Evaluating the Decency and Consistency of Data Validation Tests
Generated by LLMs\thanks{We thank Monica Alexander, Chris Maddison,
Dianne Cook, Yadong Liu, and Will Jettinghoff for helpful suggestions.
Code and data:
https://github.com/RohanAlexander/evaluating\_decency\_and\_consistency.
Contact: rohan.alexander@utoronto.ca. Contributions: RA had the initial
idea. RA and CM developed the experimental set-up. ZS established the
political donations datasets and obtained funding. LK developed the
initial suite of dataset validation tests that we compare the LLMs
against. RA wrote the code to interact with the LLMs and did the
modelling. RA, MD, and MW evaluated the LLM output. All authors
contributed to writing the initial draft as well as improving and
finalizing the paper.}}
\usepackage{etoolbox}
\makeatletter
\providecommand{\subtitle}[1]{
  \apptocmd{\@title}{\par {\large #1 \par}}{}{}
}
\makeatother
\subtitle{An application to Canadian political donations data}
\author{Rohan Alexander\footnote{Information and Statistical Sciences,
  University of Toronto. rohan.alexander@utoronto.ca.} \and Lindsay
Katz\footnote{The Investigative Journalism Foundation} \and Callandra
Moore\footnote{The Investigative Journalism Foundation} \and Michaela
Drouillard\footnote{The Investigative Journalism Foundation} \and Michael
Wing-Cheung Wong\footnote{Massachusetts Institute of Technology} \and Zane
Schwartz\footnote{The Investigative Journalism Foundation}}
\date{April 1, 2024}

\begin{document}
\maketitle
\begin{abstract}
We investigated whether large language models (LLMs) can develop data
validation tests. We considered 96 conditions each for both GPT-3.5 and
GPT-4, examining different prompt scenarios, learning modes, temperature
settings, and roles. The prompt scenarios were: 1) Asking for
expectations, 2) Asking for expectations with a given context, 3) Asking
for expectations after requesting a data simulation, and 4) Asking for
expectations with a provided data sample. The learning modes were: 1)
zero-shot, 2) one-shot, and 3) few-shot learning. We also tested four
temperature settings: 0, 0.4, 0.6, and 1. And the two distinct roles
were: 1) helpful assistant, 2) expert data scientist. To gauge
consistency, every setup was tested five times. The LLM-generated
responses were benchmarked against a gold standard data validation
suite, created by an experienced data scientist knowledgeable about the
data in question. We find there are considerable returns to the use of
few-shot learning, and that the more explicit the data setting can be
the better, to a point. The best LLM configurations complement, rather
than substitute, the gold standard results. This study underscores the
value LLMs can bring to the data cleaning and preparation stages of the
data science workflow, but highlights that they need considerable
evaluation by experienced analysts.
\end{abstract}

\section{Introduction}\label{introduction}

The Investigative Journalism Foundation (IJF) created and maintains a
dataset related to political donations in Canada. As of September 2023,
the dataset comprised 9,204,112 observations and 15 variables. Every day
new observations are added, based on newly released donations records
made available by provincial and federal elections agencies. This
release cycle is variable, with periods of inactivity followed by bursts
of multiple releases. Katz and Moore (2023) manually construct an
extensive suite of automated tests for this dataset. These impose
certain minimum standards on the dataset, including: that constituent
aspects add to match any total, class is appropriate, and null values
are as expected. This suite allows researchers to use the dataset with
confidence and ensures that new additions are fit for purpose.

We revisit this suite of tests to determine whether we can use large
language models (LLMs) to mimic this suite of validation tests. Our
estimand is the similarity between those tests written by an LLM and
those written by an experienced data scientist. We consider a variety of
prompts, roles, learning modes, and temperature settings, resulting in
an experiment with 96 conditions. In particular, we consider four prompt
variations: ask for expectations; ask for expectations given described
context; ask for expectations having first asked for a data simulation;
and ask for expectations given a sample of data. We also consider zero-,
one-, and few-shot learning; four temperature values: 0, 0.4, 0.6, and 1
(which should influence how unexpected the results are); and two roles:
helpful assistant and expert data scientist. For every combination we
obtain five responses from the LLM. We run all conditions separately for
both GPT-3.5 and GPT-4.

Three human coders judge the responses produced by the LLMs, rating
their decency (1-5, where 1 is the worst and 5 is the best) and their
consistency (1-5, where 1 means they are very different and 5 means they
are essentially identical). We then build an ordinal regression model
with \texttt{rstanarm} to explore the relationships that decency and
consistency have with prompts, roles, learning modes, and temperature
settings.

We find there are considerable returns to one- and few-shot learning. We
also find that including detailed descriptions of the expected dataset
can help improve the quality of the tests that are produced, even more
than including an example of the dataset. Temperature is associated with
consistency, but not with decency. Surprisingly, there was not much of a
difference found between GPT-3.5 and GPT-4.

Our results demonstrate one use for LLMs in production, outside of just
analysis. In particular, data scientists are often concerned that
results may be an artifact of some errors in the data cleaning and
preparation pipeline. Data validation tests help assuage these concerns
but they can be time consuming to produce. This paper shows it is
possible to use LLMs to establish an initial suite of tests, which may
encourage their use. But the variability of responses means these are
best seen as a starting place.

The remainder of this paper is structured as follows:
Section~\ref{sec-background} provides a brief overview of the underlying
dataset about which we are writing tests, as well as LLMs.
Section~\ref{sec-data} details the human coded LLM responses, which is
the analysis dataset used in this paper. Section~\ref{sec-model}
specifies the analysis model used and Section~\ref{sec-results} details
the estimates and results. Finally, Section~\ref{sec-discussion}
discusses some of the implications of this study and details a few
weaknesses.

\section{Background}\label{sec-background}

\subsection{The political donations
dataset}\label{the-political-donations-dataset}

The Investigative Journalism Foundation (IJF) is a nonprofit news media
outlet that is centered around public interest journalism. Their mandate
is to help rebuild trust in Canadian democracy and hold the powerful
accountable through data-driven investigative reporting. A core
component of the IJF's work is building and actively maintaining nine
publicly available databases with information on political donations,
registered charities, and political lobbying in Canada. While the
information that these databases are built upon are ostensibly public,
they are typically not maintained or available in a way that is widely
accessible or conducive to analysis. Further, the format and
accessibility of these data vary over time and across jurisdictions,
making it difficult to look at temporal or regional differences in the
data. The IJF's collation of these databases, and the subsequent
high-quality journalism informed by these data, serves as a crucial
contribution to rebuilding trust and transparency in Canadian democracy.

One of the IJFs eight databases, and the focus of this work, is the
political donations database. Canadian legislation requires political
parties and candidates to disclose records of financial contributions
they receive. These records are maintained by elections agencies across
provinces and territories, and at the federal level. The frequency and
scope of these disclosures vary across jurisdictions. For instance, in
British Columbia, parties, candidates, constituency associations, and
leadership and nomination contestants can receive political donations,
while in Newfoundland and Labrador, donation recipients are limited to
only parties and candidates (The IJF 2023). The IJF's political
donations database is a compilation of these political finance records
across all Canadian jurisdictions, with data spanning from 1993 to the
present day. The database contains 15 variables including the donor's
name, the political party and entity to which the donation was made, the
amount donated, as well as the region and year of the donation.

While the IJF's database is available in a clean, user-friendly format,
the original records upon which it was created were not all accessible
in this way. The format of donation records varies across jurisdiction
and time. While some are available in readily downloadable spreadsheets,
others are available as PDF or HTML files---the former necessitating the
use of optical character recognition (OCR) (The IJF 2023). To prepare
their database for publication, the IJF team performed significant
manual cleaning. The majority of this work resulted from the conversion
of PDF donation records to rectangular CSV format using OCR. This is
prone to scanning-related errors, such as the number 0 being scanned as
the letter O. The IJF manually corrected these errors wherever they were
identified by comparing the machine-legible OCR output to the original
PDF donations record (The IJF 2023).

Additional cleaning was done for the purpose of data legibility. For
instance, the IJF standardized donation dates to match the YYYY-MM-DD
format, and they standardized donor names which were in the format
``Surname, First name'' to be in the form of ``First name Surname'' (The
IJF 2023). Party names and donor types were also standardized for
consistency, and donation records with an abbreviated party name were
supplied with the complete name for that party. Finally, in rows where
the donor type was null but only individuals were legally allowed to
make donations in that jurisdiction and year, the IJF changed that null
entry to be ``Individual'' (The IJF 2023).

Despite the cleaning performed by the IJF team, the magnitude of these
data coupled with their self-reported nature makes them prone to both
human and computational error arising from the parsing process. With
over 9.2 million rows in this database, it would be difficult for the
IJF to manually identify all errors and inconsistencies present in their
data---whether that is as minor as an incorrectly formatted date, or as
major as a reported donation amount thousands of dollars above the legal
limit. To address this challenge, the IJF uses computational tools to
assess data quality (Katz and Moore 2023). The team has developed a
suite of comprehensive data tests for all nine databases using Python's
\texttt{Great\ Expectations} library. This contains a number of
pre-defined functions to test that particular characteristics expected
of a dataset hold true in the data at large.

As described by Katz and Moore (2023), the process of developing this
test suite was iterative and necessitated significant domain knowledge
to develop accurate and valuable tests. For the political donations
database, Katz and Moore (2023) developed a suite of tests pertaining to
missingness, formatting, and value expectations in the data. For
instance, they set the expectation that for donations made in British
Columbia, Ontario, or federally, the \texttt{donation\_date} entry
should not be missing, because those three jurisdictions collect that
variable (Katz and Moore 2023). They also test that, say, donations data
from 2022 align with the 2022 legal donation limits for each
jurisdiction, and that, where applicable, the sum of reported monetary
and non-monetary contribution amounts add up to the total reported
donation amount. Katz and Moore (2023) provide further details on the
development and implementation of these tests.

\subsection{Large language models}\label{large-language-models}

Historically, natural language processing (NLP) models operated in
smaller contexts than today. The first NLP systems were rule-based,
relying on hand-crafted linguistic rules based on experts' understanding
of languages, and were trained on much smaller datasets.

By the 1990s, advancement in computational capabilities and an increased
availability in larger textual datasets allowed for a paradigm shift
from rule-based systems to statistical methods. Models learned patterns
from data rather than relying on manual rules, and probabilistic models
such as Hidden Markov Models (HMMs) and Bayesian networks underpinned
new NLP applications. Where rule-based models were brittle and complex,
with existing functionalities breaking with the addition of new rules,
probabilistic models inferred linguistic patterns from large datasets,
which made them better at scaling and generalizing to solve unseen
linguistic challenges.

The transition from statistical methods and traditional machine learning
methods to ``deep learning'' came when Bengio, Ducharme, and Vincent
(2000) introduced neural network based language models. These models
predicted the next word in a sentence, given the previous words. The
model of Bengio, Ducharme, and Vincent (2000) relied on a representation
for words, where every word from the vocabulary is linked with a
continuous vector. During training, models refined these vector
representations. Words with similar meanings had vectors that were
closer together in vector space, which enabled the model to understand
semantic relationships between words.

Once neural networks were introduced, many significant advancements such
as word embeddings (Pennington, Socher, and Manning 2014; Mikolov et al.
2013), long short-term memory (Hochreiter and Schmidhuber 1997) and
Convolutional Neural Networks (CNNs) (Kim 2014) followed. The concept of
pre-training models and then fine-tuning them for specific tasks, as
seen with models like ULMFiT (Howard and Ruder 2018) and ELMo (Peters et
al. 2018), started to take root in the late 2010s, which set the stage
for transformer-based architectures.

Vaswani et al. (2017) introduced transformer architectures, which
enabled models to understand complex linguistic patterns and generate
text by processing vast amounts of text data. These models, which are
characterized by parameter counts among billions or trillions, are
trained on diverse corpora, covering cultural nuances and varieties of
human knowledge. Operating with this broader context has led to LLMs in
recent years setting new benchmarks across various NLP tasks, with a
potential to reshape many industries.

LLMs are customized and refined through in-context learning using
prompting (Ouyang et al. 2022). A prompt is a set of natural language
instructions which define the parameters and context for the desired
output and may include one or more input-output examples. By using this
approach, LLMs can apply their existing knowledge gained from training
on various datasets to adapt to new contexts specified by the prompt.
The outcomes generated by LLMs are sensitive to the specific phrasing
and structure of these natural language instructions. Consequently,
there is ongoing work to establish effective prompt patterns (White et
al. 2023).

There has been some work using LLMs for data cleaning, including H.
Zhang et al. (2023) who focus on data pre-processing.

\section{Data}\label{sec-data}

We are interested in the extent to which the LLMs can develop a suite of
data validation tests that is similar to a suite developed by an
experienced expert data scientist who is familiar with the dataset. To
test this, we establish and run a series of experiments where we
consider different specifications and then compare the LLM output. The
variables that we consider are summarized here, and the full details are
included in Appendix~\ref{sec-details}.

\begin{itemize}
\tightlist
\item
  Four prompts:

  \begin{itemize}
  \tightlist
  \item
    List the variables of interest---``index'', ``amount'',
    ``donor\_location'', ``donation\_date'', ``donor\_full\_name'',
    ``donor\_type'', ``political\_entity'', ``political\_party'',
    ``recipient'', ``region'', ``donation\_year'', ``amount\_monetary'',
    ``amount\_non\_monetary'', ``electoral\_event'',
    ``electoral\_district'', ``added''---and ask for a suite of
    expectations using the Python package \texttt{great\_expectations}.
  \item
    As above, plus provide details about each variable. For instance,
    ``amount'' is a monetary value that cannot be less than \$0. An
    example observation is ``195.46''. It is, possible, but unlikely to
    be more than \$1,000.00. It cannot be NA. It should be a numeric.
    The maximum donation ``amount'' depends on the value of ``region''
    and ``year''. For ``Federal'' is 1675, for ``Quebec'' is 100 since
    2013 and 500 for earlier years, for ``British Columbia'' is 1309.09,
    for ``Ontario'' is 3325, and for ``Alberta'' is 4300. There is no
    limit for ``Saskatchewan''.
  \item
    As above, plus ask the LLM to simulate an example dataset of 1000
    observations.
  \item
    Provide the details of the dataset in the first option, as well as
    an extract of the dataset.
  \end{itemize}
\item
  Three learning modes: zero-, one-, and few-shot learning. In the
  first, we do not provide any examples; in the second, we provide one
  example; and in the third, we provide three examples.
\item
  Four temperature values: 0, 0.4, 0.6, and 1.
\item
  Two roles:

  \begin{itemize}
  \tightlist
  \item
    You are a helpful assistant.
  \item
    You are a highly-trained, experienced, data scientist who is an
    expert at writing readable, correct, Python code.
  \end{itemize}
\end{itemize}

This combination of variables and options results in 96 different prompt
situations. We run these through both GPT-3.5 and GPT-4, using the
OpenAI API. For every combination we ask for five responses to
understand variation.

This results in a dataset of responses. The combination of settings that
gave rise to each response was blinded and the order randomized, and
then the responses were ranked by three experienced human evaluators on
two metrics.

\begin{enumerate}
\def\labelenumi{\arabic{enumi}.}
\tightlist
\item
  Human evaluators ranked the ``decency'' of the first response for each
  combination of variables. This is a measure of how effective the LLM
  validation tests were compared with the code written by the
  experienced data scientist who wrote the original suite of tests. The
  LLM responses are not expected to have the full context of the code,
  so we do not expect an exact match, but it should actually write code,
  import relevant libraries, add comments, deal with class, and write a
  variety of relevant expectations. 1 means that the code is unusable, 2
  means that it is not unusable but would need a lot of work and would
  be disappointing from a human, 3 means that it is fine but would need
  some fixing, 4 means it is broadly equivalent to what the gold
  standard suite contains, and 5 means it is in no worse and is better
  in some way than the gold standard validation suite. The decency
  rating is at a condition level, focusing on only the first response
  given that combination of settings. We consider only the first given
  constraints on the evaluators' time.
\item
  The other, ``consistency'', is a ranking 1-5 of how different each of
  the five responses was for that particular combination of variables. 1
  means that responses 1-5 were wildly different. 5 means that responses
  1-5 are entirely or essentially the same.
\end{enumerate}

\begin{figure}

\centering{

\includegraphics{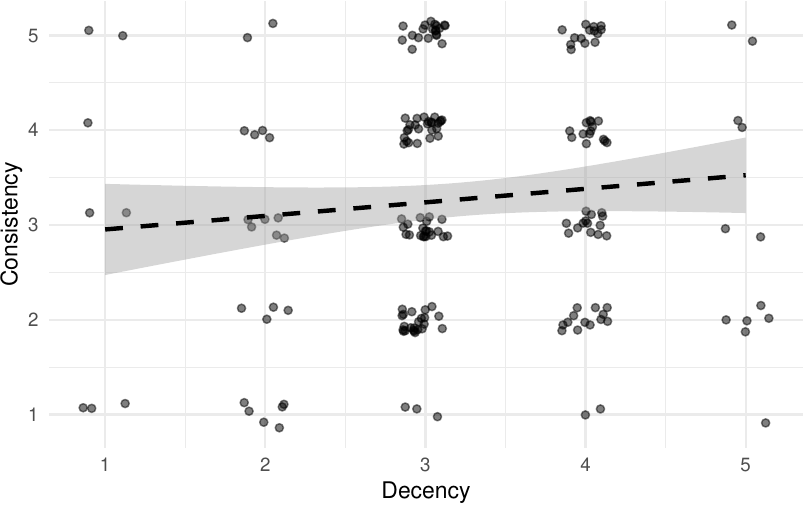}

}

\caption{\label{fig-decencyvsconsistency}How decency and consistency
compare with each other. Higher decency means the LLM responses
correspond more closely with the gold standard. Higher consistency means
all five responses were similar to each other.}

\end{figure}

To bring the three evaluators together we consider the median response.
This then allows us to examine whether there is a relationship between
decency and consistency (Figure~\ref{fig-decencyvsconsistency}). There
is a broadly positive relationship. Details about coder-specific
responses are provided in Appendix~\ref{sec-intercoder}.

\begin{figure}

\begin{minipage}{0.50\linewidth}

\centering{

\includegraphics{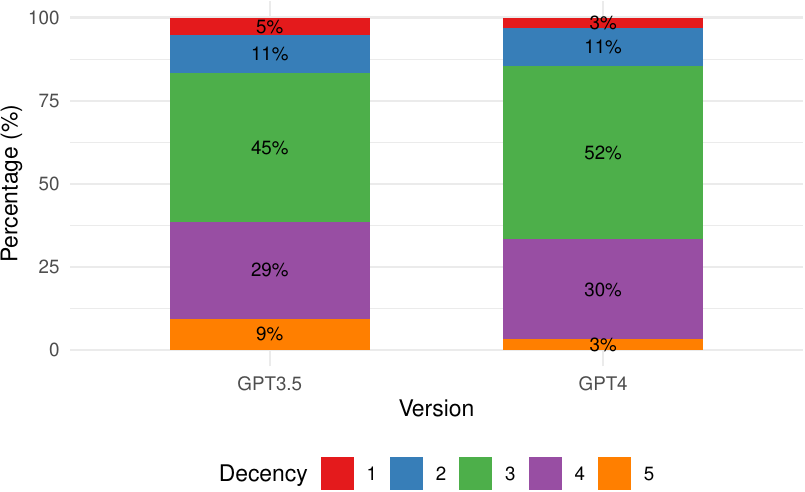}

}

\subcaption{\label{fig-version-1}Decency}

\end{minipage}%
\begin{minipage}{0.50\linewidth}

\centering{

\includegraphics{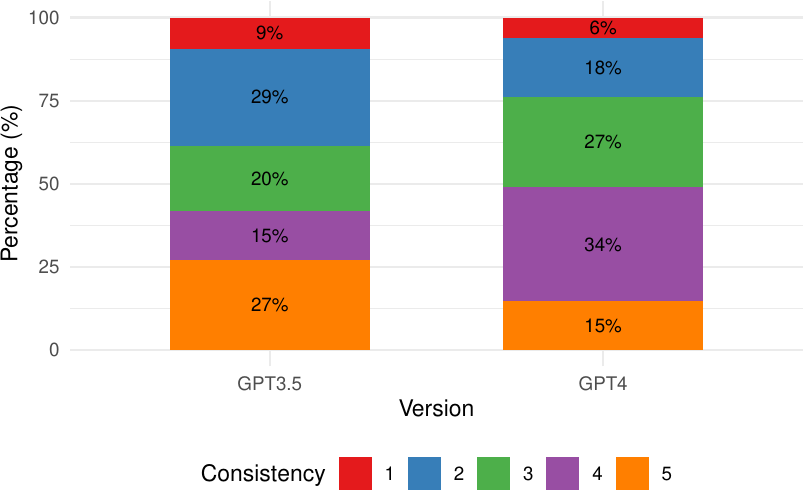}

}

\subcaption{\label{fig-version-2}Consistency}

\end{minipage}%

\caption{\label{fig-version}How decency and consistency change based on
whether GPT-3.5 or GPT-4 is used. For decency, 1 is the worst and 5 is
the best. For consistency 5 means all responses were essentially the
same, while 1 means at least one was quite different.}

\end{figure}

Figure~\ref{fig-version} illustrates how decency and consistency differ
based on whether GPT-3.5 or GPT-4 is used. Unexpectedly, GPT-4 has fewer
responses rated 5/5 for decency, compared with GPT-3.5 with 3\% compared
with 9\% (Figure~\ref{fig-version-1}). Overall, the mean decency
response for GPT-3.5 is 3.26, while the mean decency of the responses
generated by GPT-4 is 3.19. The consistency is not too different between
the two versions, with GPT-3.5 having an average of 3.21, while GPT-4
has an average of 3.33. Unexpectedly, GPT-4 appears to have fewer
responses that are essentially identical i.e.~5/5
(Figure~\ref{fig-version-2}).

\begin{figure}

\begin{minipage}{0.50\linewidth}

\centering{

\includegraphics{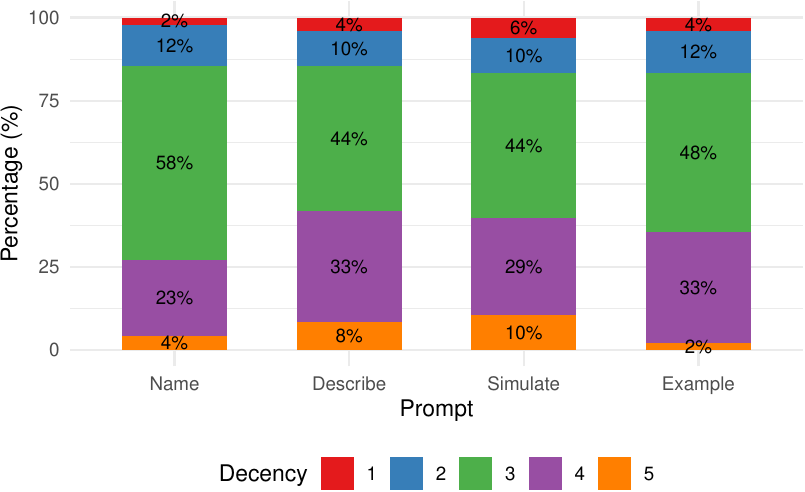}

}

\subcaption{\label{fig-prompt-1}Decency}

\end{minipage}%
\begin{minipage}{0.50\linewidth}

\centering{

\includegraphics{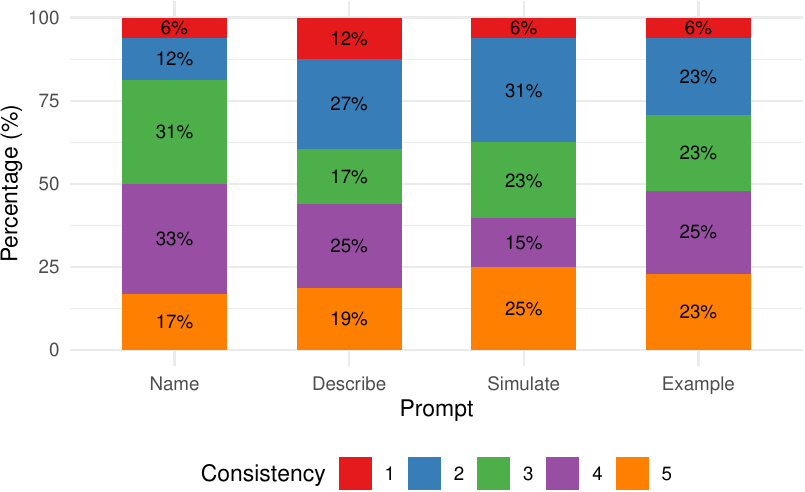}

}

\subcaption{\label{fig-prompt-2}Consistency}

\end{minipage}%

\caption{\label{fig-prompt}How decency and consistency change based on
the type of prompt used. Name means essentially only the variables of
interest are listed. Describe means some detail about each variable is
provided. Simulate means the LLM is asked to first think about
simulating the variables. Example means an extract of the dataset is
provided.}

\end{figure}

Figure~\ref{fig-prompt} examines how decency and consistency differ
based on which prompt is used. The prompts differ by how much
information is provided. The least informative prompt, ``Name'',
essentially consists of just providing the LLM with the names of the
columns that are expected to be in the dataset. The next most
informative prompt, ``Describe'', adds a detailed description of what we
expect of the observations. ``Simulate'' adds that we expect the LLM to
first simulate a dataset based on that description, before generating
the expectations. And finally, the most informative prompt, ``Example'',
provides a snapshot of the dataset, consisting of the relevant variables
and ten observations.

There is a difference in terms of how the prompts are associated with
decency (Figure~\ref{fig-prompt-1}). In particular, ``Name'' is almost
never associated with a 5/5 rating, and it has many responses rated 3/5.
Surprisingly the most informative prompt, ``Example'', is also never
associated with a 5/5 rating. Instead it is ``Describe'' and
``Simulate'' that tend to be associated with better decency ratings.
This is reflected in the averages, which are 3.15, 3.31, 3.27, and 3.17
for ``Name'', ``Describe'', ``Simulate'', and ``Example'', respectively.

The pattern is less clear when it comes to consistency
(Figure~\ref{fig-prompt-2}). All four have similar averages, at 3.42,
3.10, 3.21, and 3.35 for ``Name'', ``Describe'', ``Simulate'', and
``Example'', respectively. A wider variety of responses (as denoted by
lower consistency ratings), is rarely seen for ``Name'' and
``Describe''. It is surprising that ``Name'' should have the highest
average.

Temperature is a parameter that varies from 0 to 1, that we use to
manipulate how random a LLM is. At high temperatures, an LLM will
produce a wider variety of responses. At lower temperatures it will
focus on the single most likely response. Higher temperature should be
associated with a wider variety of LLM responses.

\begin{figure}

\begin{minipage}{0.50\linewidth}

\centering{

\includegraphics{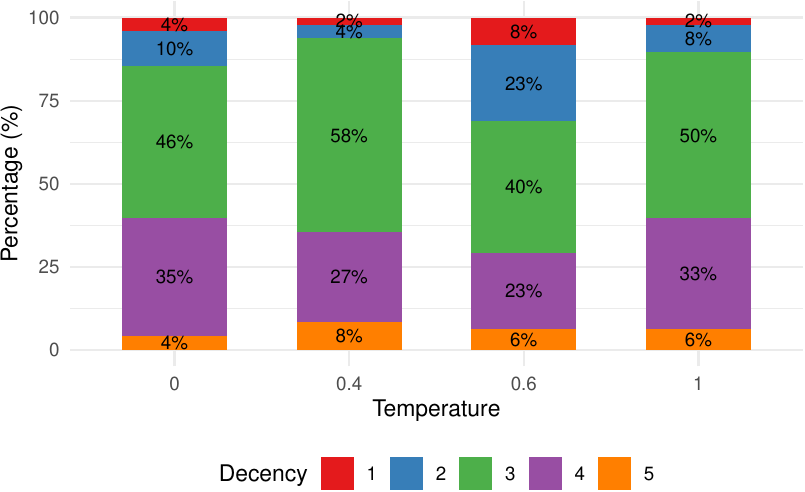}

}

\subcaption{\label{fig-temperature-1}Decency}

\end{minipage}%
\begin{minipage}{0.50\linewidth}

\centering{

\includegraphics{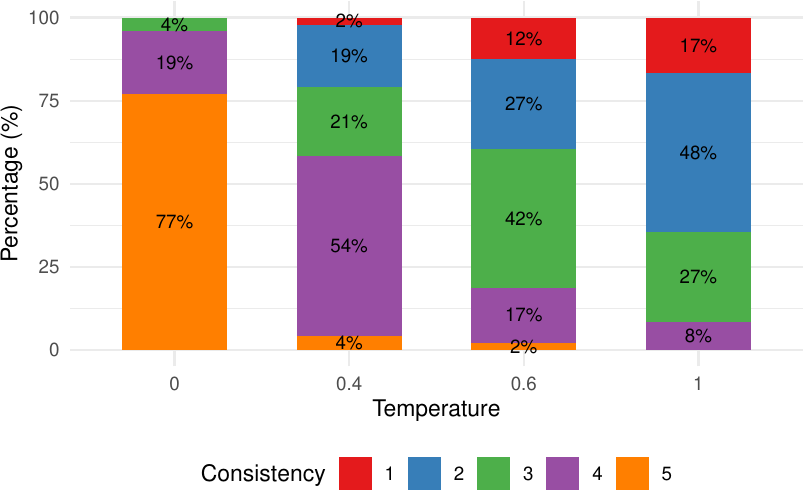}

}

\subcaption{\label{fig-temperature-2}Consistency}

\end{minipage}%

\caption{\label{fig-temperature}How decency and consistency change based
on temperature. Temperature of 0 means that less variation is expected
from the outputs, while temperature of 1 means considerable variation is
expected.}

\end{figure}

Figure~\ref{fig-temperature} examines how decency and consistency differ
based on which of the four temperature values we consider---0, 0.4, 0.6,
1---is used. There appears to be limited difference in terms of how
different temperature values are associated with decency
(Figure~\ref{fig-temperature-1}). They all have similar mean values at
3.25, 3.35, 2.96, and 3.33 for temperature values of 0, 0.4, 0.6, and 1,
respectively.

In contrast, a sizeable different in consistency can be seen with
different temperatures values (Figure~\ref{fig-temperature-2}).
Temperature values of 0 are associated with ratings of high consistency,
and higher temperatures are associated with lower consistency. Their
means differ considerably, with 4.73, 3.40, 2.69, and 2.27 for
temperature values of 0, 0.4, 0.6, and 1, respectively. The one
unexpected aspect is temperature of 0.6, which has an outsized number of
responses with decency of 1/5 or 2/5.

\begin{figure}

\begin{minipage}{0.50\linewidth}

\centering{

\includegraphics{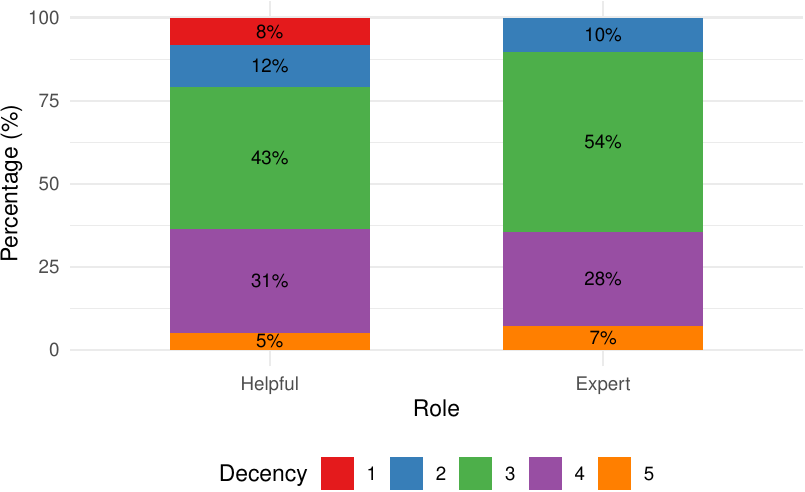}

}

\subcaption{\label{fig-role-1}Decency}

\end{minipage}%
\begin{minipage}{0.50\linewidth}

\centering{

\includegraphics{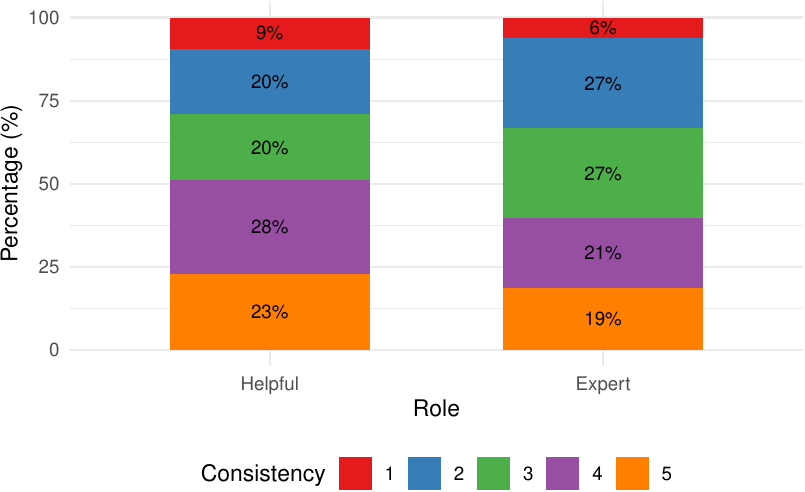}

}

\subcaption{\label{fig-role-2}Consistency}

\end{minipage}%

\caption{\label{fig-role}How decency and consistency change based on
whether role is `Helpful' or `Expert'. This is considered before the
main prompt content and is designed to position the LLM as having a
certain role.}

\end{figure}

Role is an aspect of a prompt that is provided to the LLM before the
main prompt content. We considered two different roles, one that
positioned the LLM as a helpful assistant, and the other that positioned
the LLM as an experienced data scientist (Figure~\ref{fig-role}). We
were expecting that the expert role would result in better code, but
there was no obvious difference in terms of decency
(Figure~\ref{fig-role-1}) or consistency (Figure~\ref{fig-role-2}).
Although it is notable that there are no responses rated 1/5 with the
``Expert'' role. Their mean decency did not differ by much in either
case with 3.12 and 3.32 for ``Helpful'' and ``Expert'', respectively,
while mean consistency was 3.35 and 3.19.

Learning mode refers to the number of examples provided to the LLM as
part of the prompt. Zero-shot learning means that no examples are
provided, while one-shot and few-shot learning refer to one- and a few-
examples being provided, respectively. Although the advantage of LLMs
such as GPT-3.5 and GPT-4 is that they can do well with zero-shot
learning, we would expect that they will do better with one-shot and
few-shot.

\begin{figure}

\begin{minipage}{0.50\linewidth}

\centering{

\includegraphics{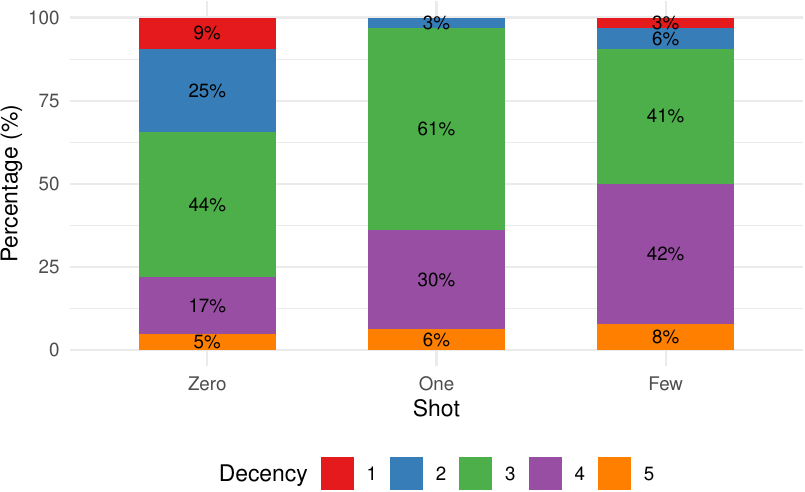}

}

\subcaption{\label{fig-shot-1}Decency}

\end{minipage}%
\begin{minipage}{0.50\linewidth}

\centering{

\includegraphics{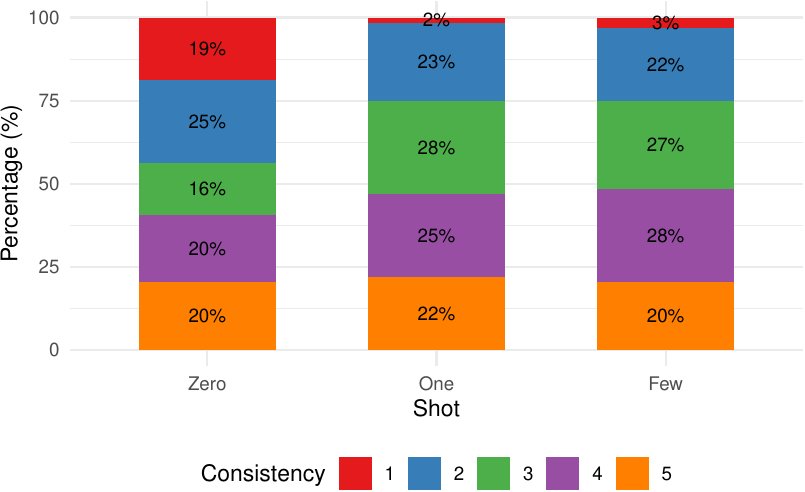}

}

\subcaption{\label{fig-shot-2}Consistency}

\end{minipage}%

\caption{\label{fig-shot}How decency and consistency change based on
zero-, one-, and few-shot learning. This refers to the number of
examples given to the LLM with zero-shot learning meaning no examples
are provided, while few-shot learning means a few are given.}

\end{figure}

We see substantial differences, especially when moving away from
zero-shot learning (Figure~\ref{fig-shot}). In particular, we see that
decency of 1/5 and 2/5 is dominated by zero-shot
(Figure~\ref{fig-shot-1}). This is also reflected in the mean decency
which for zero-shot is 2.83, while for one- and few-shot learning is
3.39 and 3.45, respectively. This is largely due to the relative
reduction in 1/5 and 2/5.

We see this pattern in consistency as well. For instance, zero-shot
learning is over-represented in the least consistent responses, both 1/5
and 2/5 (Figure~\ref{fig-shot-2}). And the mean level of consistency is
lower for zero-shot learning, at 2.98, compared with single- and
few-shot, at 3.42 and 3.41, respectively.

\section{Model}\label{sec-model}

Consistency and decency, our dependent variables, are ordered,
categorical, outcomes. As such we model the rating \(Y\), which is an
ordinal outcome with \(J = 5\) possible categories, using ordered
logistic regression. Such a model works by estimating the probability
that the outcome is less than or equal to some specific category, which
in this case is the coded rank of that response (i.e.~1, 2, 3, 4, 5).
Models for consistency and decency are estimated separately. In latent
variable formulation, we assume that the observed category \(y\) can be
related to an underlying continuous latent variable, \(y^*\), through a
series of cutpoints:

\[
y=\left\{\begin{array}{ll}
1 & \text { if } y^*<\zeta_1 \\
2 & \text { if } \zeta_1 \leq y^* \\
\vdots & \\
J & \text { if } \zeta_{I-1}<1
\end{array}\right.
\]

where \(\zeta\) is a vector of cutpoints. The latent variable \(y^*\) is
then assumed to have a logistic distribution and is modeled as a linear
function of covariates. In our case, we are interested in exploring the
relationships that consistency and decency have with model, prompt,
temperature, role, and learning mode:

\[
y^* = \beta_1 \cdot \mbox{version}_i + \beta_2 \cdot \mbox{prompt}_i + \beta_3 \cdot \mbox{temperature}_i + \beta_4 \cdot \mbox{role}_i + \beta_5 \cdot \mbox{shot}_i
\]

In terms of our predictors, version is a binary variable as to whether
we are using GPT-3.5 or GPT-4. Before looking at the data, we expected
that GPT-4 would do better in terms of both decency and consistency than
GPT-3.5. Prompt can be one of four values: ``Name'', ``Describe'',
``Simulate'', and ``Example''. We expected that ``Simulate'' and
``Example'' would be associated with higher decency than ``Name'' and
``Describe''. However, we expected the opposite relationship with
consistency. This is because we expected that what would increase
decency would be the more specific guidance provided by the ``Simulate''
and ``Example'' prompts, which would also increase the consistency.
Temperature can take one of four values: 0, 0.4, 0.6 and 1. We expected
that higher temperature values would be associated with less
consistency. However, it was difficult to anticipate how temperature
should be related to decency. It could be that allowing more
``randomness'' enables better responses, or it could be that it allows
more scope to be wrong. Role can be one of two values: ``Helpful'', or
``Expert'', while learning mode can be one of three: ``zero'', ``one'',
or ``few''. In the case of both role and learning mode, we expected that
the ``Expert'' role, and one- and few-shot learning would be associated
with higher decency, and consistency.

We fit this model, separately, for each of consistency and decency, in a
Bayesian framework using the package \texttt{rstanarm} (Goodrich et al.
2023) and the R statistical programming language (R Core Team 2023).
This computational process requires priors to be placed on \(R^2\), the
proportion of variance in the outcome that is attributable to the
coefficients in a linear model, and the vector of cutpoints \(\zeta\).
For the prior on \(R^2\), we follow Gelman, Hill, and Vehtari (2020,
276) and assume the mean is 0.3. For the vector cutpoints \(\zeta\), we
use an uninformative prior of \(\mbox{Dirichlet}(1)\). Gabry and
Goodrich (2020) provide more information about fitting this type of
model using \texttt{rstanarm}. Model diagnostics are provided in
Appendix~\ref{sec-diagnostics}.

\section{Results}\label{sec-results}

\subsection{Model-based results}\label{model-based-results}

\begin{figure}

\centering{

\includegraphics{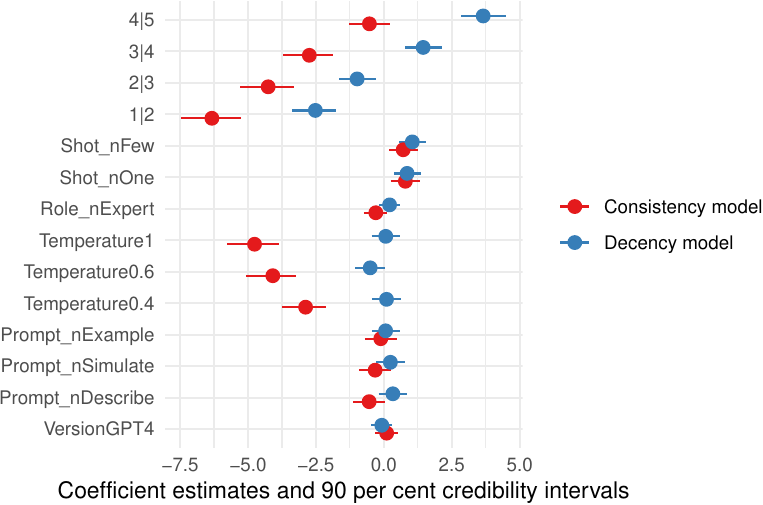}

}

\caption{\label{fig-modelsummaryestimates}Exploring the relationships
that consistency and decency have with model, prompt, temperature, role,
and learning mode. The x-axis shows coefficient estimates and 90 per
cent credibility intervals.}

\end{figure}

In total we considered 192 observations, which is 96 for each GPT-3.5
and GPT-4. The estimates from our models are shown in
Table~\ref{tbl-modelsummaryestimates} and
Figure~\ref{fig-modelsummaryestimates}, which were both produced with
\texttt{modelsummary} (Arel-Bundock 2022).

\begin{table}

\caption{\label{tbl-modelsummaryestimates}Exploring the relationships
that consistency and decency have with model, prompt, temperature, role,
and learning mode. The mad (mean absolute deviation) statistic is in
brackets.}

\centering{

\centering
\begin{tabular}[t]{lcc}
\toprule
  & Consistency model & Decency model\\
\midrule
VersionGPT4 & \num{0.11} & \num{-0.07}\\
 & (\num{0.25}) & (\num{0.23})\\
Prompt\_nDescribe & \num{-0.54} & \num{0.33}\\
 & (\num{0.36}) & (\num{0.31})\\
Prompt\_nSimulate & \num{-0.32} & \num{0.24}\\
 & (\num{0.35}) & (\num{0.33})\\
Prompt\_nExample & \num{-0.11} & \num{0.07}\\
 & (\num{0.35}) & (\num{0.32})\\
Temperature0.4 & \num{-2.88} & \num{0.10}\\
 & (\num{0.49}) & (\num{0.33})\\
Temperature0.6 & \num{-4.09} & \num{-0.50}\\
 & (\num{0.55}) & (\num{0.33})\\
Temperature1 & \num{-4.76} & \num{0.07}\\
 & (\num{0.58}) & (\num{0.32})\\
Role\_nExpert & \num{-0.30} & \num{0.21}\\
 & (\num{0.26}) & (\num{0.24})\\
Shot\_nOne & \num{0.79} & \num{0.86}\\
 & (\num{0.32}) & (\num{0.29})\\
Shot\_nFew & \num{0.71} & \num{1.05}\\
 & (\num{0.32}) & (\num{0.30})\\
1|2 & \num{-6.33} & \num{-2.52}\\
 & (\num{0.65}) & (\num{0.49})\\
2|3 & \num{-4.26} & \num{-0.98}\\
 & (\num{0.60}) & (\num{0.42})\\
3|4 & \num{-2.74} & \num{1.45}\\
 & (\num{0.56}) & (\num{0.42})\\
4|5 & \num{-0.53} & \num{3.66}\\
 & (\num{0.47}) & (\num{0.49})\\
\midrule
Num.Obs. & \num{192} & \num{192}\\
ELPD & \num{-224.7} & \num{-240.3}\\
ELPD s.e. & \num{8.1} & \num{10.0}\\
LOOIC & \num{449.5} & \num{480.6}\\
LOOIC s.e. & \num{16.2} & \num{20.1}\\
WAIC & \num{449.4} & \num{480.6}\\
\bottomrule
\end{tabular}

}

\end{table}%

Table~\ref{tbl-modelsummaryestimates} shows the coefficient estimate
with the mad (mean absolute deviation) statistic in brackets. In
Figure~\ref{fig-modelsummaryestimates}, the dot represents the
coefficient estimate, while the lines are 90 per cent credibility
intervals.

The intercepts, 1\textbar2, 2\textbar3, 3\textbar4, and 4\textbar5,
shown in Table~\ref{tbl-modelsummaryestimates} reflect cutpoints, where
a rating goes from one category to the next (e.g.~from 1 to 2, etc).

The models do not identify a substantial difference between GPT-3.5 and
GPT-4 in terms of consistency, and, surprisingly, even a somewhat
negative association in terms of decency.

They also do not identify much difference between the different prompt
types in terms of consistency. However, they do find that ``Describe''
and ``Simulate'' are associated with increased decency, as is
``Example'', although surprisingly to a lesser extent.

As expected, increases in temperature are strongly associated with less
consistency. However, we are unable to identify much of a relationship
between temperature and decency.

The models struggle to find any association between which role is used
in terms of either consistency or decency. That is, priming the prompt
with either ``You are a helpful assistant.'' or ``You are a
highly-trained, experienced, data scientist who is an expert at writing
readable, correct, Python code.'' did not seem to change much in terms
of the coding of the responses.

Finally, one- and few-shot learning is associated with increased decency
and slightly increased consistency over zero-shot prompts. That said,
the magnitude of this relationship is not overly large.

\subsection{Coder-based results}\label{coder-based-results}

The rating of the coherence and decency was a human endeavor, with all
the bias that brings. One benefit is that the coders were able to leave
comments about the LLM outputs as they rated them. The provides an
opportunity for additional results, in addition to those based on their
quantitative ratings.

In terms of decency, sometimes the LLMs simply did not generate code,
instead generating natural language text about what tests could look
like. Other times the LLMs generated a lot of very similar tests, for
instance going through many possible options testing for class or
non-null results, at the expense of developing a broadly useful suite.
Finally, sometimes the LLMs would generate only one or two tests, or
generate tests without any comments or other explanations.

The responses that ``excelled'' in their decency scores tended to bring
a new tool beyond what was in the gold standard code suite -- for
instance, generating a class with functions, or saving the results as a
JSON.

No test written by an LLM surpassed the gold standard suite in terms of
testing for something that was not tested for in the human written test
suite. Responses that wrote multiple tests for data types or to check
that the response is not null were ranked worse than responses with a
richer variety of tests. The highest quality tests that the LLMs
generated were conditional tests (for instance, checking for a value in
a column if the value of another column is X), which are also included
in the gold standard suite.

With regard to consistency, it was often the case that a low consistency
rating was due to one of the five responses being especially different.
It was rare that the LLM just entirely misinterpreted the prompt, but
compared with a human data scientist, who brings a broad context to
writing tests, the LLMs rarely took into account the entire context that
we would like.

In some cases, one or two results would generate sentences explaining
which expectations to test for instead of code, while the other results
generated commented code. It was also often the case that results would
test for similar things, but one result would write the code slightly
different (for instance, using an if statement or a function while
testing for the same thing).

The results that scored highest in consistency were exactly identical,
or extremely similar with slight variations in comments or tests.

\section{Discussion}\label{sec-discussion}

\subsection{On the importance and challenges of writing validation
tests}\label{on-the-importance-and-challenges-of-writing-validation-tests}

Data validation is a cornerstone of high quality data science workflows.
Broadly, data validation is centered around establishing the credibility
of individual data values, the coherence of data with itself, and the
consistency of data with relevant external sources (Alexander 2023).
Data quality---or a lack thereof---can have fundamental downstream
effects on the rest of the scientific process. If the data being
supplied to a statistical model is incorrect in some way (e.g., a
categorical variable is stored as numeric), this will have significant
implications for subsequent results and conclusions (Hynes, Sculley, and
Terry 2017; Breck et al. 2019; L. Zhang et al. 2023; Gao, Xie, and Tao
2016).

One way to implement data validation in practice is automated testing,
which intuits data expectations from format, column name, or other meta
information and converts those expectations to code (L. Zhang et al.
2023). Automated validation promotes trustworthiness and transparency.
By developing a suite of tests to validate that particular
characteristics of a dataset hold true, data scientists uncover
fundamental assumptions that underpin their work and equip users with
confidence that those assumptions are met in practice. The value of data
validation testing is not limited to particular domains---it is a
fundamental component of a reliable data science workflow, regardless of
the data at hand (Alexander 2023; Gao, Xie, and Tao 2016).

There are a number of challenges that accompany the development and
implementation of data validation tests. These challenges are
particularly important when thinking about data validation efforts
across research domains and levels of statistical and data science
expertise.

Implementing automated tests requires the ability to develop code. There
are a number of testing-centered libraries in various programming
languages. \texttt{Great\ Expectations}---the Python library employed by
Katz and Moore (2023)---contains various data testing functions and has
a data assistant functionality which programmatically explores data to
develop tests based on observed characteristics (Great Expectations
2023). Similarly, \texttt{pointblank} is an R package which offers
predefined automated data testing functionality (Iannone and Vargas
2023). With a focus on data validation for machine learning, Hynes,
Sculley, and Terry (2017) developed a data linter tool that detects
possible inconsistencies in the data and suggests appropriate
transformations for a specific model type. Further, Breck et al. (2019)
executed an automated validation machine learning platform at Google to
address concerns about the downstream effects of messy data in model
training. While various data testing software have been developed and
shared, many of these have limited functionality, and the implementation
of a test suite may require significant data processing work and
necessarily requires programming knowledge (Alexander 2023; L. Zhang et
al. 2023).

From their work on data validation for the IJF Katz and Moore (2023)
conclude that data tests should not be compromised just to be compatible
with predefined test functions that exist in a software package. In
order to develop the most valuable suite of tests, it may be necessary
to create new variables in a dataset, merge multiple datasets, or create
tests which are outside the scope of existing software frameworks, for
example. This presents a challenge for individuals who are not trained
in or comfortable with computer programming, but wish to
programmatically test their data.

In addition, the development of a comprehensive, bespoke, and accurate
data test suite can rarely be done by one person alone. As advocated by
Katz and Moore (2023), domain expertise is fundamental to the creation
and development of high-quality data tests. For instance, without
incorporating the knowledge that donation dates are only collected in
three jurisdictions (British Columbia, Ontario and Federal) into their
code, Katz and Moore (2023)'s test for missingness for that variable
would have produced inflated failure rates. This is knowledge that was
gained through collaborative efforts with IJF team members who were
involved in dataset construction. For an individual working on an
unfamiliar dataset, seeking out domain knowledge and gaining a thorough
understanding of that data is a non-trivial task in and of itself.

The process of designing, developing, and deploying data validation
tests is iterative in nature and takes a great deal of time (Katz and
Moore 2023). For researchers or journalists who are eager to publish
their findings, for instance, there may not be enough incentive or
resources for them to do this time-consuming validation work. This
challenge is only amplified if an individual was not involved in the
dataset construction process, does not have access to vital domain
expertise, or does not have programming experience.

While data validation is valuable for the production of transparent and
trustworthy research, it comes with many challenges which hinder its
accessibility. This work presents an opportunity to advocate for an
alternative approach to data validation through the use of LLMs. Though
there has been a great deal of work done to develop more accessible
programming tools for data validation, these tools do not entirely
mitigate the challenges which accompany the process, namely the need for
programming knowledge, domain expertise, and a great deal of time. Our
findings illustrate the potential for LLMs to rapidly produce usable
data validation tests, written in code, with limited information---a
promising application to support the promotion of more trustworthy,
transparent data workflows across domains.

\subsection{On the use of LLMs in statistics and data
science}\label{on-the-use-of-llms-in-statistics-and-data-science}

In this paper we find considerable improvements with the use of few-shot
learning and being more explicit about the data that we are interested
in having tests for. That said, the LLM outputs that were considered are
a starting place for an experienced data scientist, rather than a
replacement for the data scientist.

The contributions of this experiment and this paper are twofold.
Firstly, this work adds to the emerging area of inquiry of prompt
engineering for large language models. In doing so, our prompts may
serve as examples---both good and bad---for others hoping to implement
LLMs into their research and data workflows. Secondly, this work
contributes to the academic study of LLMs for the development of data
science methodologies. As these tools become more powerful and the
precision in our ability to specify their outputs improves, they will
become tools to reduce the time and resource burden of writing software
and tests.

Our work is consistent with previous literature demonstrating that LLM
performance on complex, user-defined tasks is sensitive to prompt
engineering. Prompting is a brittle process in which minor alterations
can lead to large fluctuations in performance (Arora et al. 2022; Zhao
et al. 2021), however with OpenAI's LLMs in particular, GPT-4 (Giorgi et
al. 2023) seems less susceptible to this than GPT-3.5. In general,
previous research has demonstrated that when using GPT-3.5, prompts
should provide context (White et al. 2023; Clavi\' et al. 2023), be
specific (White et al. 2023; Yong et al. 2023; Shieh, n.d.), define a
persona or role (White et al. 2023; Clavi\'{e} et al. 2023), break complex
reasoning into smaller steps (White et al. 2023; Henrickson and
Meroño-Peñuela 2023), and provide example responses (Brown et al. 2020;
Arora et al. 2022; Shieh, n.d.) to improve performance on a variety of
tasks, including classifying job types (Clavi\'{e} et al. 2023), generating
images for construction defect detection (Yong et al. 2023), and
generating hermeneutically valuable text (Henrickson and Meroño-Peñuela
2023).

These observations are somewhat borne out in our experiment. For
instance, the more specified prompt (``Describe'') and the prompt
eliciting step-by-step thinking (``Simulate'') both show improvements
over the simplest prompt (``Name''). However, improvements made by
specificity break down when example data is provided (``Example''). We
hypothesize that this is due to GPT-3.5 and GPT-4's poor performance on
numerical and structured data in comparison to more domain-specific LLMs
such as BloombergGPT which has been trained on numeric, structured
financial data and utilizes a tokenizer more suited to these data (Wu et
al. 2023). We do not see significant changes in performance when two
different roles are specified to the LLM, which contradicts previous
literature (White et al. 2023; Clavi\'{e} et al. 2023). Finally, our
significant improvements moving from zero- to one-shot and marginal to
no improvements from one- to few-shot align with existing literature
(Giorgi et al. 2023). Some studies have demonstrated GPT-3.5 and GPT-4's
difficulties in responding to queries in the desired format during
zero-shot structured tasks, including named entity recognition (Hu et
al. 2023) and multiple choice question (Clavi\'{e} et al. 2023). Giorgi et
al. (2023) thus attribute performance improvements in one- and few-shot
prompts to the in-context examples' guidance as to the desired output
structure. Upon inspection of the outputs from zero-, one-, and few-shot
prompts, we believe that this same effect exists for our prompts.

Our work additionally contributes to the growing body of work using LLMs
to automate portions of the data science workflow. LLMs have already
been integrated into software tools such as GitHub's Co-Pilot ({``{Your
AI pair programmer},''} n.d.) and AutoGPT ({``{AutoGPT},''} n.d.) and
IDEs such as IntelliJ (Krochmalski 2104) and VSCode (Dias 2023) to
produce code in a variety of programming languages. These tools will
continue to assist data scientists to accelerate, resolve bugs in,
document, and optimize, their code in all parts of the data science
workflow. LLMs have also demonstrated potential to automate other
onerous or tedious tasks in data science pipelines. These include
generating synthetic text data (Chung, Kamar, and Amershi 2023; Yu et
al. 2023), improving search-based software testing (Lemieux et al.
2023), generating unit tests from natural language in a variety of
programming languages and contexts (Lahiri et al. 2023; Schäfer et al.
2023), generating property-based tests from API documentation (Vikram,
Lemieux, and Padhye 2023), and conducting exploratory data analysis (Ma
et al. 2023). As LLM prompting and fine-tuning methods improve, they
will reduce technical barriers to conducting transparent, accurate,
reproducible analysis, enabling data scientists to be more productive
and put their energy towards analysis. higher-order reasoning, and
sophisticated inference. The work described in this paper towards
automating data validation using LLMs has the capacity to further this
evolution.

\subsection{Limitations}\label{limitations}

There are a variety of limitations of this study. The most notable is
that there were only three evaluators of the LLMs responses. It is
possible that some of the evaluation, especially that for decency, would
have benefited from additional evaluators. The study also only considers
one, reasonably specific, setting. Although it is likely this is
reasonably representative of many messy real-world datasets, expanding
our approach to consider a variety of settings could be beneficial.

A limitation of our work on prompt engineering (and indeed of prompt
engineering generally) is that we are unable to provide fully reliable
explanations for improvements in performance. Though we can make
convincing speculations, these are ultimately educated guesses. Previous
studies have shown that in addition to being brittle to precise
semantics (Arora et al. 2022; Zhao et al. 2021), prompts show improved
performance from additional text which provides no new information
(Henrickson and Meroño-Peñuela 2023) such as naming the model (e.g.,
``You are Frederick, a helpful AI''), emulating the chatbot saying that
it understands the instructions, asking it to output the ``right
conclusion'', and providing positive feedback (Clavi\'{e} et al. 2023).
Despite the opacity of the emergent properties of LLMs, we may be able
to develop greater understanding of the impacts of prompting choices by
testing these prompts on a greater diversity of datasets from across
disciplines and contexts.

We were focused on writing tests in Python within the
\texttt{Great\ Expectations} framework. It may be that testing these
prompts in a variety of contexts, and also considering potential avenues
of exploration beyond \texttt{Great\ Expectations}, will provide
evidence as to whether our prompting strategies are generalizable across
datasets in different formats, contexts, and technical regimes. Prompts
can be conceptualized as knowledge transfer methods analogous to
software patterns which are intended to provide reusable solutions to
common problems (White et al. 2023). This aligns with the goal of
automated data validation to reduce the need for domain expertise and
resources dedicated to data validation. Thus, in order to improve the
explainability of our methods and to ensure our method is generalizable,
we must validate that the prompts that are effective on this dataset
show similar performance on other datasets from different contexts, in
different formats, and with other LLMs.

\newpage

\appendix

\section*{Appendix}\label{appendix}
\addcontentsline{toc}{section}{Appendix}

\section{Prompt details}\label{sec-details}

\begin{itemize}
\item
  Four prompts:

  \begin{lstlisting}
      - The Investigative Journalism Foundation (IJF) created and maintains a CSV dataset related to political donations in Canada. Each observation in the dataset is a donation, and the dataset has the following variables: "index", "amount", "donor_location", "donation_date", "donor_full_name", "donor_type", "political_entity", "political_party", "recipient", "region", "donation_year", "amount_monetary", "amount_non_monetary", "electoral_event", "electoral_district", "added".\n\rPlease write a series of expectations using the Python package great_expectations for this dataset.
      - The Investigative Journalism Foundation (IJF) created and maintains a CSV dataset related to political donations in Canada. Each observation in the dataset is a donation, and the dataset has the following variables: "index", "amount", "donor_location", "donation_date", "donor_full_name", "donor_type", "political_entity", "political_party", "recipient", "region", "donation_year", "amount_monetary", "amount_non_monetary", "electoral_event", "electoral_district", "added". 
          - "amount" is a monetary value that cannot be less than $0. An example observation is "195.46". It is, possible, but unlikely to be more than $1,000.00. It cannot be NA. It should be a numeric. The maximum donation "amount" depends on the value of "region" and "year". For "Federal" is 1675, for "Quebec" is 100 since 2013 and 500 for earlier years, for "British Columbia" is 1309.09, for "Ontario" is 3325, and for "Alberta" is 4300. There is no limit for "Saskatchewan".
          - "amount" should be equal to the sum of "amount_monetary" and "amount_non_monetary".
          - "region" can be one of the following values: "Federal", "Quebec", "British Columbia", "Ontario", "Saskatchewan", "Alberta". It cannot be NA. It should be a factor variable.
          - "donor_full_name" is a string. It cannot be NA. It is usually a first and last name, but might also include a middle initial. It should be in title case.
          - "donation_date" should be a date in the following format: YYYY-MM-DD. It could be NA. The earliest donation is from 2010-01-01. The latest donation is from 2023-09-01.
          - "donation_year" should match the year of "donation_date" if "donation_date" is not NA, but it is possible that "donation_year" exists even if "donation_date" does not. The earliest year is 2010 and the latest year is 2023. This variable is an integer.
          - "political_party" cannot be NA. It should be a factor that is equal to one of: "New Democratic Party", "Liberal Party of Canada", "Conservative Party of Canada".

          Please write a series of expectations using the Python package great_expectations for this dataset.
      - The Investigative Journalism Foundation (IJF) created and maintains a CSV dataset related to political donations in Canada. Each observation in the dataset is a donation, and the dataset has the following variables: "index", "amount", "donor_location", "donation_date", "donor_full_name", "donor_type", "political_entity", "political_party", "recipient", "region", "donation_year", "amount_monetary", "amount_non_monetary", "electoral_event", "electoral_district", "added". 
          - "amount" is a monetary value that cannot be less than $0. An example observation is "195.46". It is, possible, but unlikely to be more than $1,000.00. It cannot be NA. It should be a numeric. The maximum donation "amount" depends on the value of "region" and "year". For "Federal" is 1675, for "Quebec" is 100 since 2013 and 500 for earlier years, for "British Columbia" is 1309.09, for "Ontario" is 3325, and for "Alberta" is 4300. There is no limit for "Saskatchewan".
          - "amount" should be equal to the sum of "amount_monetary" and "amount_non_monetary".
          - "region" can be one of the following values: "Federal", "Quebec", "British Columbia", "Ontario", "Saskatchewan", "Alberta". It cannot be NA. It should be a factor variable.
          - "donor_full_name" is a string. It cannot be NA. It is usually a first and last name, but might also include a middle initial. It should be in title case.
          - "donation_date" should be a date in the following format: YYYY-MM-DD. It could be NA. The earliest donation is from 2010-01-01. The latest donation is from 2023-09-01.
          - "donation_year" should match the year of "donation_date" if "donation_date" is not NA, but it is possible that "donation_year" exists even if "donation_date" does not. The earliest year is 2010 and the latest year is 2023. This variable is an integer.
          - "political_party" cannot be NA. It should be a factor that is equal to one of: "New Democratic Party", "Liberal Party of Canada", "Conservative Party of Canada".

          Please simulate an example dataset of 1000 observations. Based on that simulation please write a series of expectations using the Python package great_expectations for this dataset.
      - The Investigative Journalism Foundation (IJF) created and maintains a CSV dataset related to political donations in Canada. Each observation in the dataset is a donation, and the dataset has the following variables: "index", "amount", "donor_location", "donation_date", "donor_full_name", "donor_type", "political_entity", "political_party", "recipient", "region", "donation_year", "amount_monetary", "amount_non_monetary", "electoral_event", "electoral_district", "added". 
          An example of a dataset is: 
          index,amount,donor_location,donation_date,donor_full_name,donor_type,political_entity,political_party,recipient,region,donation_year,amount_monetary,amount_non_monetary,electoral_event,electoral_district,added
          5279105,$20.00,"Granton, N0M1V0",2014-08-15,Shelley Reynolds,Individual,Party,New Democratic Party,New Democratic Party,Federal,2014,20.0,0.0,Annual,Nan,2022-11-23 01:00:31.771769+00:00
          2187800,$200.00,,,Robert Toupin,Individual,Party,Coalition Avenir Qu\'{e}bec - l'\'{e}quipe Fran\c{c}ois Legault,,Quebec,2018,,,,,2023-03-17 18:02:29.706250+00:00
          3165665,$50.00,,,Genevi\'{e}ve Dussault,Individual,Party,Qu\'{e}bec Solidaire  (Avant Fusion),,Quebec,2017,,,,,2023-03-19 18:02:24.746621+00:00
          8803473,$250.00,"Nan, Nan",,Roger Anderson,Individual,Party,Reform Party Of Canada,Reform Party Of Canada,Federal,1994,0.0,0.0,Annual,Nan,2022-11-22 02:25:34.868056+00:00
          2000776,"$1,425.00","Calgary, T3H5K2",2018-10-30,Melinda Parker,Individual,Registered associations,Liberal Party Of Canada,Calgary Centre Federal Liberal Association,Federal,2018,1425.0,0.0,Annual,Calgary Centre,2022-11-23 01:00:31.771769+00:00
          9321613,$75.00,,2022-06-17,Jeffrey Andrus,Individual,Party,Bc Ndp,Bc Ndp,British Columbia,2022,,,,,2022-12-21 02:20:49.009276+00:00
          2426288,$50.00,"Stony Plain, T7Z1L5",2018-07-24,Phillip L Poulin,Individual,Party,Conservative Party Of Canada,Conservative Party Of Canada,Federal,2018,50.0,0.0,Annual,Nan,2022-11-23 01:00:31.771769+00:00
          4428629,$100.00,"Calgary, T2Y4K1",2015-07-30,Barry Hollowell,Individual,Party,New Democratic Party,New Democratic Party,Federal,2015,100.0,0.0,Annual,Nan,2022-11-23 01:00:31.771769+00:00
          1010544,$20.00,"Langley, V1M1P2",2020-05-31,Carole Sundin,Individual,Party,New Democratic Party,New Democratic Party,Federal,2020,20.0,0.0,Annual,Nan,2022-11-23 01:00:31.771769+00:00
          4254927,$500.00,"Welshpool, E5E1Z1",2015-10-10,Melville E Young,Individual,Party,Conservative Party Of Canada,Conservative Party Of Canada,Federal,2015,500.0,0.0,Annual,Nan,2022-11-23 01:00:31.771769+00:00
          8001740,$90.00,"Deleau, R0M0L0",2004-11-15,Clarke Robson,Individual,Party,New Democratic Party,New Democratic Party,Federal,2004,90.0,0.0,Annual,Nan,2022-11-23 01:00:31.771769+00:00

          Based on this sample please write a series of expectations using the Python package great_expectations for this dataset.
  \end{lstlisting}
\item
  Three learning modes: zero-, one-, and few-shot learning:

  \begin{lstlisting}
      - The following text in quotes is an example of an expectation for this dataset:
          """
          # Check that there is nothing null in any column of donations details
          donations_mv.expect_column_values_to_not_be_null(column='donor_full_name')
          """
      - The following text in quotes is an example of three expectations for this dataset:
          """
          # Check that there is nothing null in any column of donations details
          donations_mv.expect_column_values_to_not_be_null(column='donor_full_name')
          # Check that the federal donation does not exceed the maximum
          donations_mv.expect_column_values_to_be_between(
              column = 'amount',
              max_value = 1675,
              row_condition = 'region=="Federal" & donor_full_name.str.contains("Contributions Of")==False & donor_full_name.str.contains("Estate Of")==False & donor_full_name.str.contains("Total Anonymous Contributions")==False & donation_year == 2022 & political_entity.str.contains("Leadership")==False',
              condition_parser = 'pandas'
          )
          # Check that the date matches an appropriate regex format
          donations_mv.expect_column_values_to_match_regex(column = 'donation_date',
                                                      regex = '\\d{4}-\\d{2}-\\d{2}',
                                                      row_condition = "donation_date.isna()==False",
                                                      condition_parser = 'pandas')
          """
  \end{lstlisting}
\item
  Four temperature values: 0, 0.4, 0.6, and 1.
\item
  Two roles:

  \begin{lstlisting}
      - You are a helpful assistant.
      - You are a highly-trained, experienced, data scientist who is an expert at writing readable, correct, Python code.
  \end{lstlisting}
\end{itemize}

\newpage

\section{Intercoder differences}\label{sec-intercoder}

\begin{figure}

\begin{minipage}{0.50\linewidth}

\centering{

\includegraphics{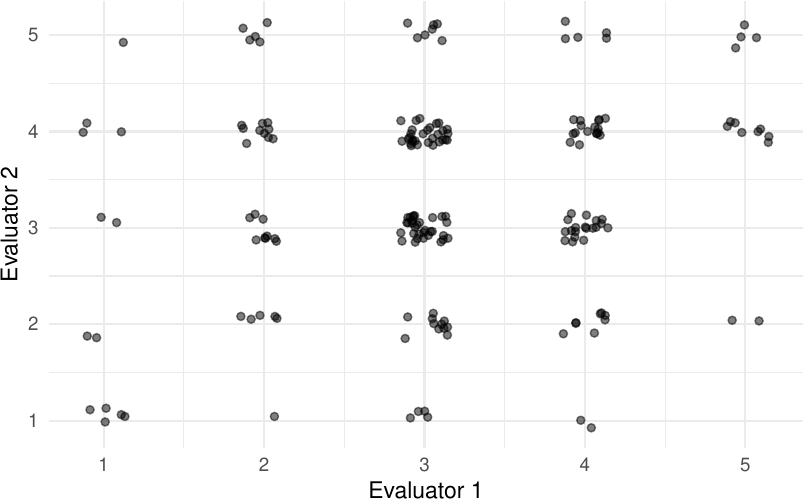}

}

\subcaption{\label{fig-intercoder-1}Decency}

\end{minipage}%
\begin{minipage}{0.50\linewidth}

\centering{

\includegraphics{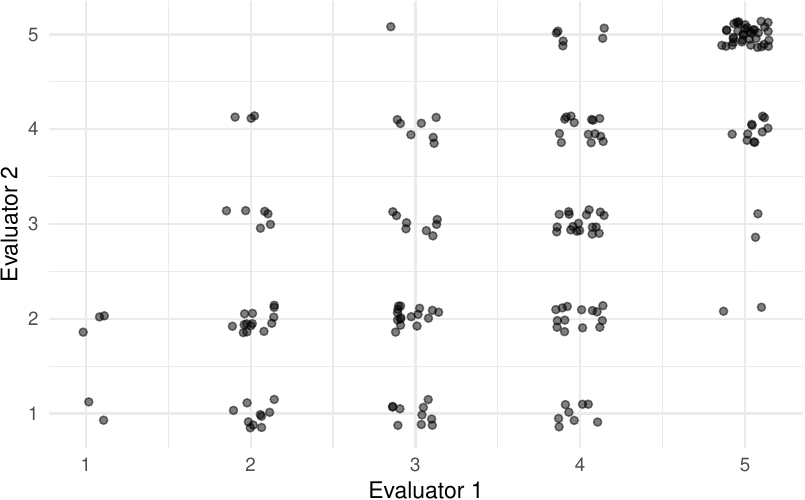}

}

\subcaption{\label{fig-intercoder-2}Consistency}

\end{minipage}%
\newline
\begin{minipage}{0.50\linewidth}

\centering{

\includegraphics{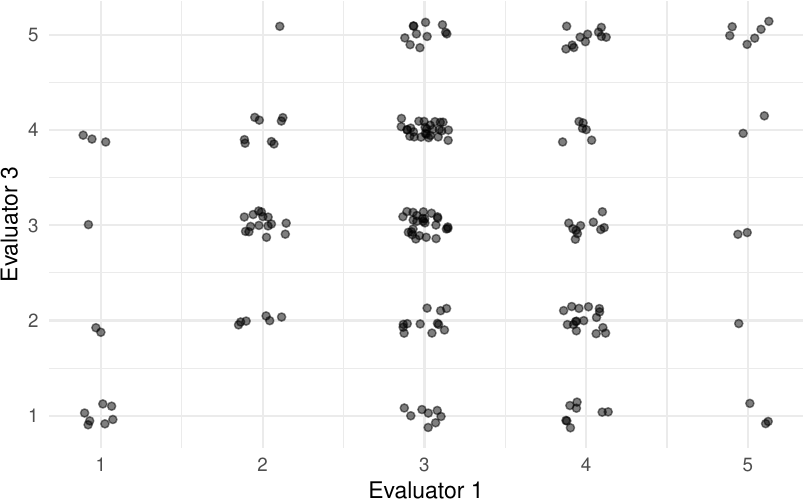}

}

\subcaption{\label{fig-intercoder-3}Decency}

\end{minipage}%
\begin{minipage}{0.50\linewidth}

\centering{

\includegraphics{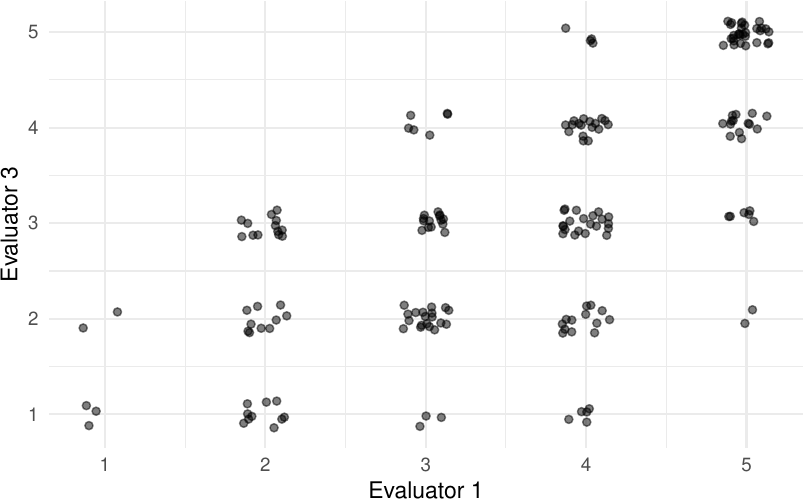}

}

\subcaption{\label{fig-intercoder-4}Consistency}

\end{minipage}%
\newline
\begin{minipage}{0.50\linewidth}

\centering{

\includegraphics{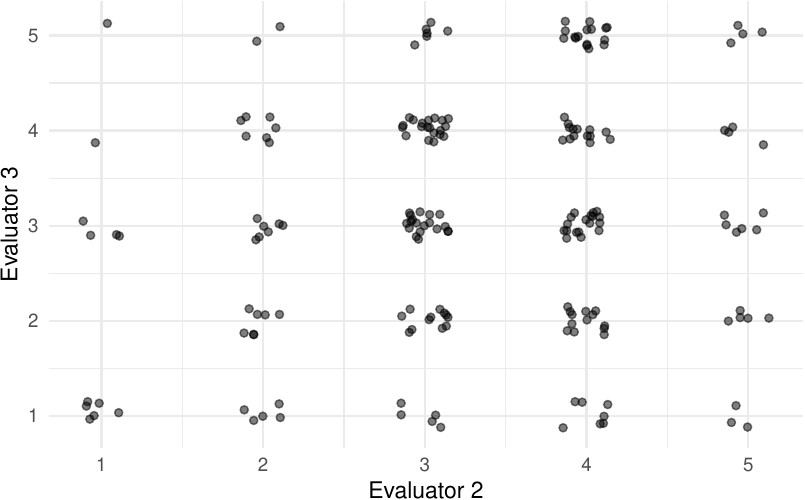}

}

\subcaption{\label{fig-intercoder-5}Decency}

\end{minipage}%
\begin{minipage}{0.50\linewidth}

\centering{

\includegraphics{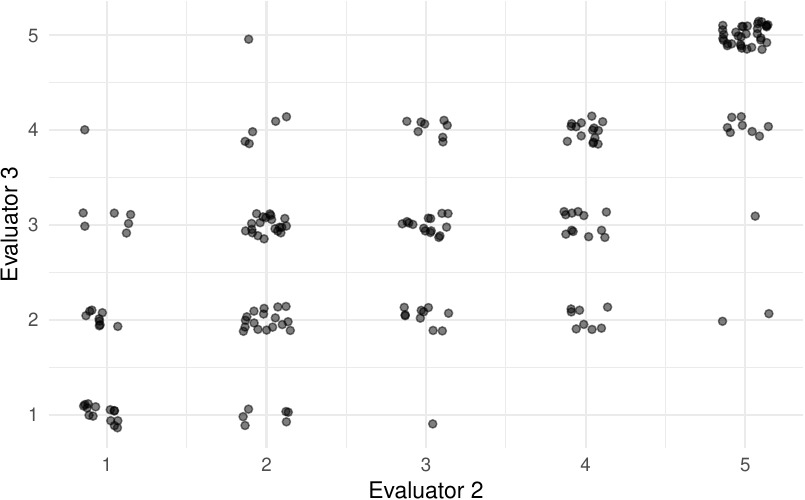}

}

\subcaption{\label{fig-intercoder-6}Consistency}

\end{minipage}%

\caption{\label{fig-intercoder}How decency and consistency are coded by
each evaluator}

\end{figure}

Figure~\ref{fig-intercoder} examines how the three evaluators rated
decency and consistency. In particular, Figure~\ref{fig-intercoder-1}
focuses on how they agreed on decency, and Figure~\ref{fig-intercoder-2}
shows how they agreed on consistency. If all evaluators rated every
response the same, then we would expect all responses to be on the 45
degree line. We do not find many absolute differences of opinion, but
there nonetheless is still a wide range of differences, especially in
terms of decency.

\newpage

\section{Model diagnostics}\label{sec-diagnostics}

\subsection{Consistency model}\label{consistency-model}

\begin{figure}[H]

{\centering \includegraphics{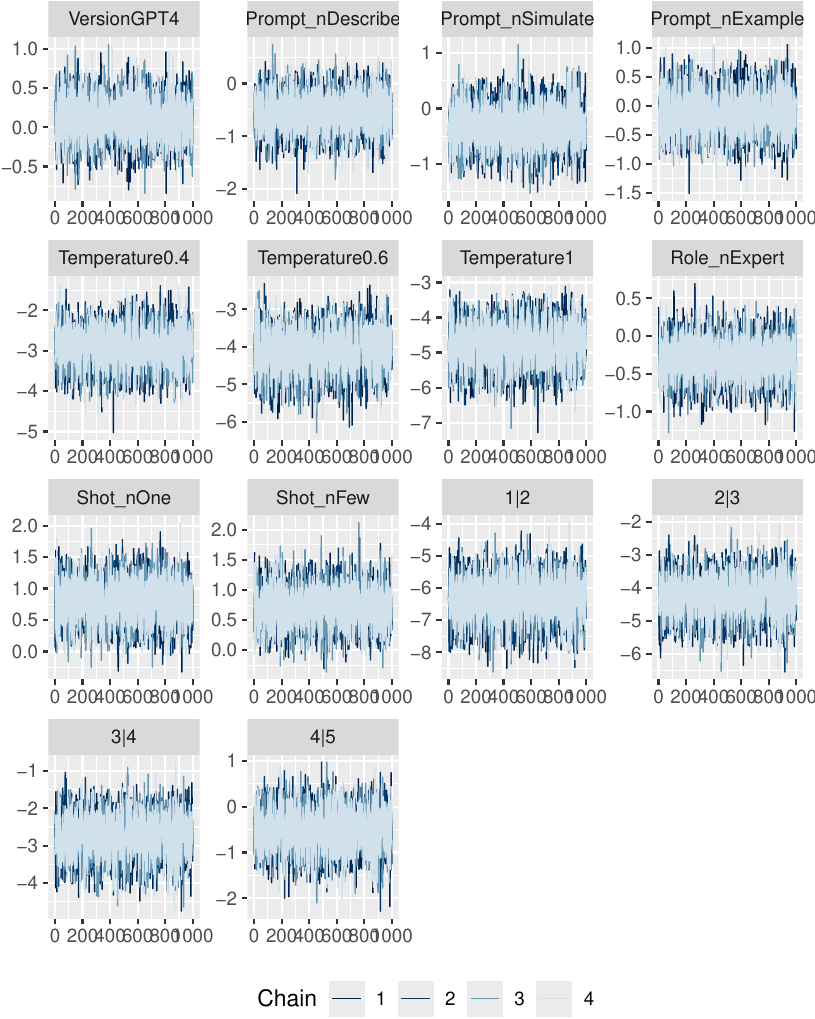}

}

\caption{Consistency model trace plot}

\end{figure}%

\begin{figure}[H]

{\centering \includegraphics{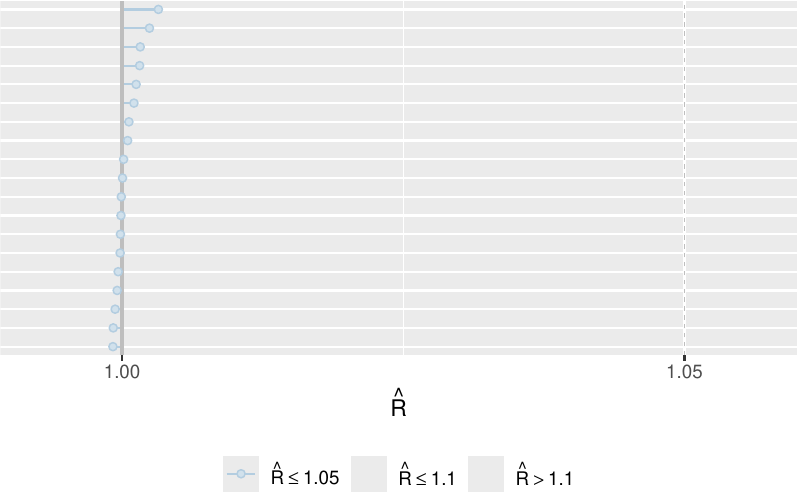}

}

\caption{Consistency model rhat plot}

\end{figure}%

\begin{figure}[H]

{\centering \includegraphics{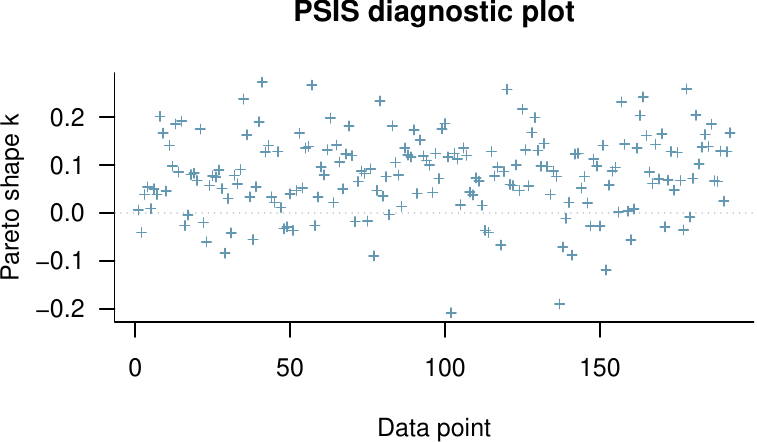}

}

\caption{Consistency model loo plot}

\end{figure}%

\begin{figure}[H]

{\centering \includegraphics{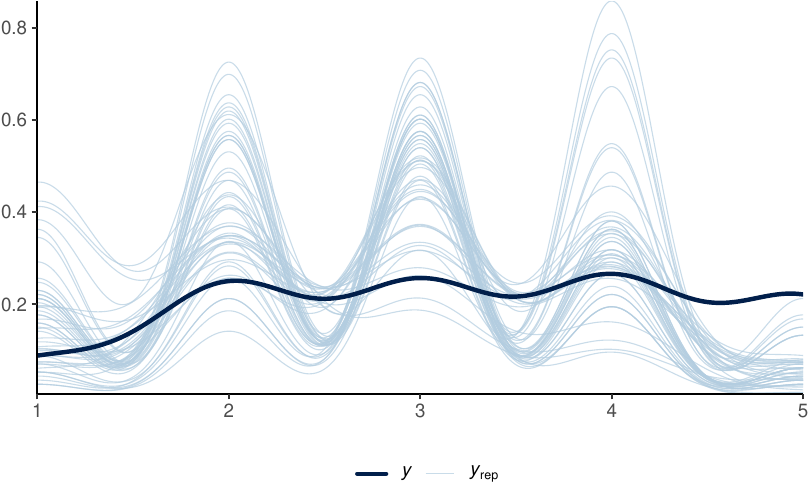}

}

\caption{Consistency model posterior predictive check}

\end{figure}%

\begin{figure}[H]

{\centering \includegraphics{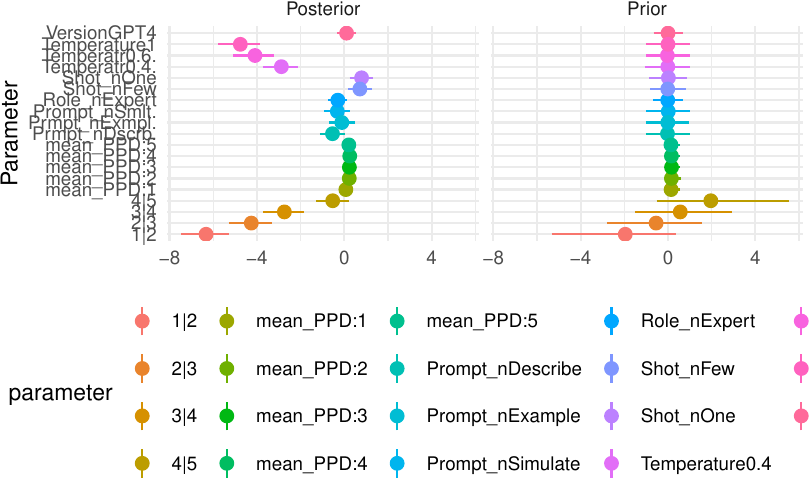}

}

\caption{Consistency model posterior vs prior}

\end{figure}%

\newpage

\subsection{Decency model}\label{decency-model}

\begin{figure}[H]

{\centering \includegraphics{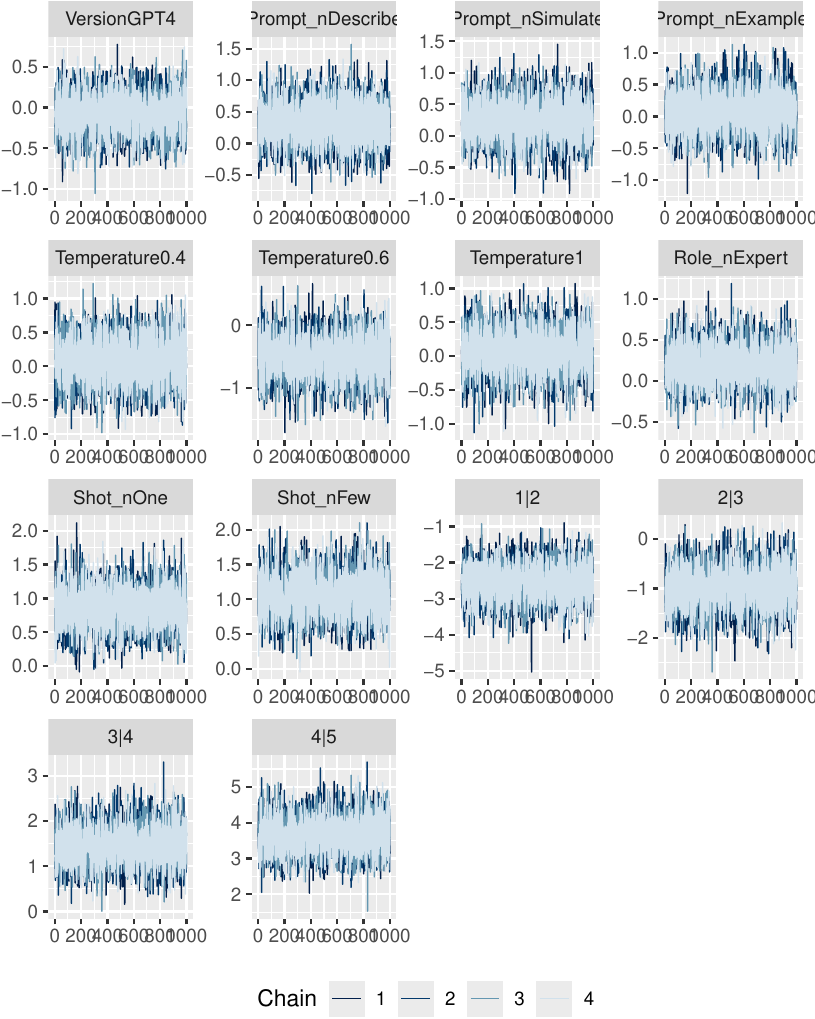}

}

\caption{Decency model trace plot}

\end{figure}%

\begin{figure}[H]

{\centering \includegraphics{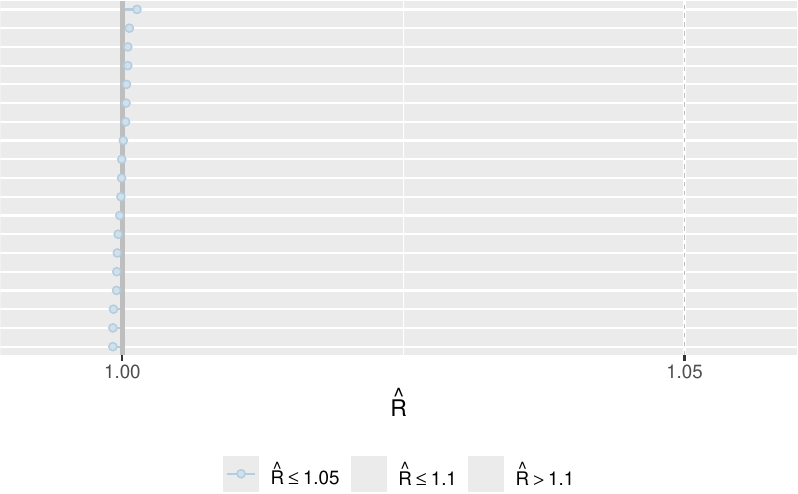}

}

\caption{Decency model rhat plot}

\end{figure}%

\begin{figure}[H]

{\centering \includegraphics{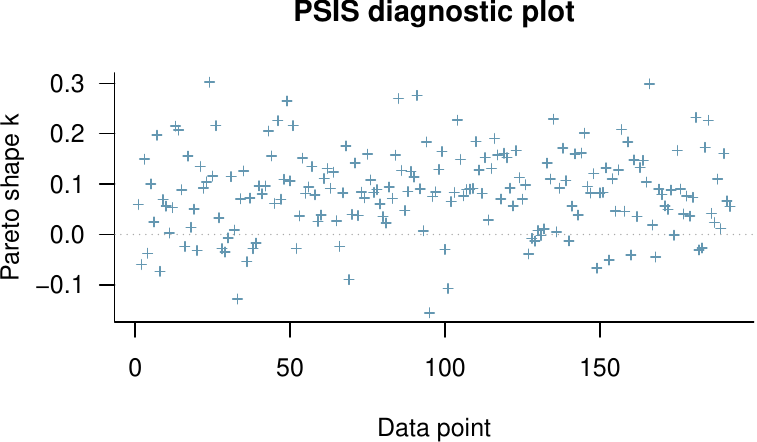}

}

\caption{Decency model loo plot}

\end{figure}%

\begin{figure}[H]

{\centering \includegraphics{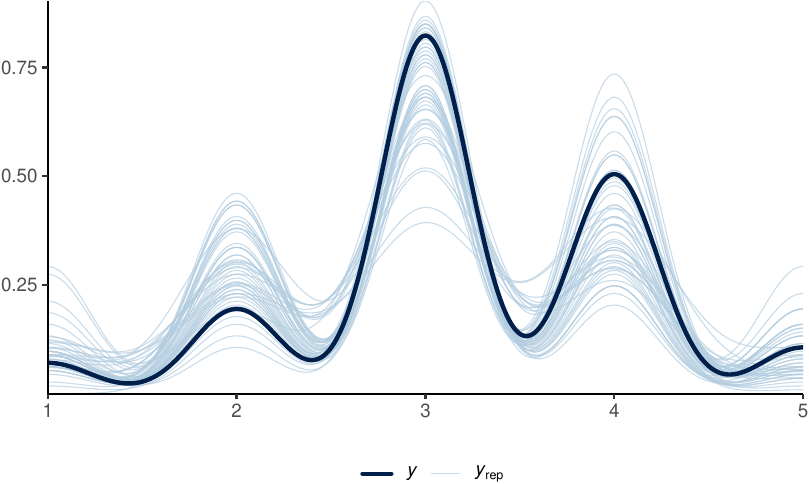}

}

\caption{Decency model posterior predictive check}

\end{figure}%

\begin{figure}[H]

{\centering \includegraphics{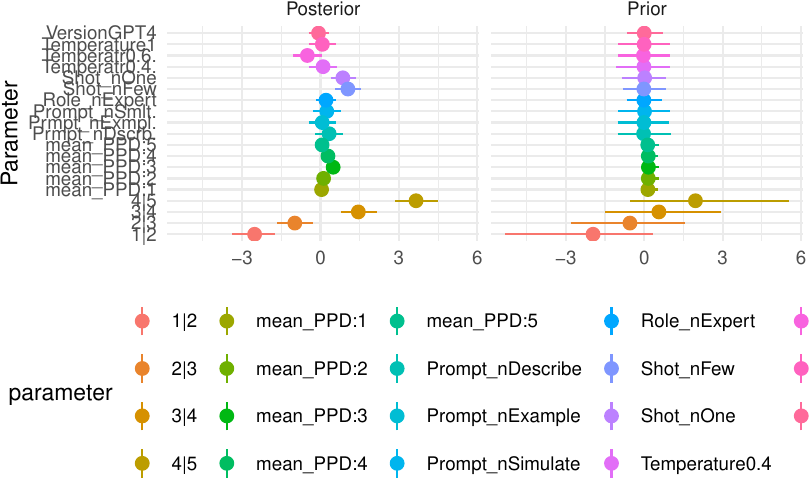}

}

\caption{Decency model posterior vs prior}

\end{figure}%

\newpage

\section*{References}\label{references}
\addcontentsline{toc}{section}{References}

\phantomsection\label{refs}

\end{document}